\documentclass[12pt]{article}

\usepackage{epsfig}
\usepackage{amsmath}
\usepackage{verbatim}
\usepackage{bbm}

\begin{document}

\begin{flushright}
{\tt MAN/HEP/2006/35}
\end{flushright}

\begin{center}

{\Large\bf Radiative Lifting of Flat Directions of the MSSM\\
\vskip 4mm
in de Sitter Background}

\vskip 1cm

Bj\"orn Garbrecht\\
{\it
School of Physics \& Astronomy,
The University of~Manchester,\\
Oxford~Road,
Manchester M13~9PL, United Kingdom}

\end{center}

\vskip .5cm

\begin{abstract}
The one-loop effective potential for a charged scalar field in de Sitter
background is studied. We derive an approximate form for the gauge boson
propagator with a general covariant gauge-fixing parameter $\xi$. This
expression is used to derive an explicitly gauge-invariant
effective potential,
which agrees with earlier results obtained in Landau gauge.
Adding the scalar- and fermion-loop contributions, that arise in a
supersymmetric Higgs model, we find that the effective potential is vanishing
to first order in the Hubble rate $H^2$. The contribution due to a chiral
multiplet running in the loop is however non-vanishing. When applied to
a flat direction of the MSSM with VEV $w$, a logarithmically
running correction of order $h^2 H^2 w^2$ arises, where $h $ is a Yukawa
coupling. This is of particular importance for $D$-term inflation or
models endowed with a Heisenberg symmetry, where Hubble scale mass terms
for flat directions from supergravity corrections are absent.

\end{abstract}

\section{Introduction}

The scalar potential landscape
of the Minimal Supersymmetric Standard Model (MSSM)
is endowed with a large number
of flat directions~\cite{Gherghetta:1995dv}. These lead to a
variety of possible
cosmological scenarios~\cite{Enqvist:2003gh}, the most prominent
of which is the Affleck-Dine mechanism for baryogenesis~\cite{Affleck:1984fy}.
Recently, also possible effects on the thermal history of the Universe have
gained attention~\cite{Allahverdi:2005fq,Olive:2006uw}.
Of course, flat directions are not utterly flat, and
for both, qualitative and quantitative predictions in cosmology,
a detailed knowledge of their lifting is necessary. In the present day
Universe, the flatness is lifted by renormalizable soft supersymmetry (SUSY)
breaking operators of TeV scale, which are constrained by
particle physics phenomenology, as well as by non-renormalizable operators,
which become important when the vacuum expectation value (VEV) along the
flat direction approaches the superstring scale. The non-renormalizable
terms are in general unknown, although some restraints can be gained
from the supergravity scalar
potential~\cite{Copeland:1994vg,Dine:1995kz,Dine:1995uk}.
In addition, one can also derive the minimum dimension of a
lifting operator for a given flat direction~\cite{Gherghetta:1995dv}.

During inflation, where the initial conditions for the subsequent
evolution of the VEVs are set, there are further possibilities for
uplifting. We mention here three important ways.
The first is spontaneous breakdown of SUSY
by the vacuum energy density driving
inflation~\cite{Garbrecht:2006az}. In $F$-term
models~\cite{Copeland:1994vg,Dvali:1994ms},
the MSSM particle content generically couples to the non-zero $F$-term,
which drives inflation, {\it via} a Grand Unified symmetry breaking Higgs field
or even through the Electroweak two-Higgs multiplet~\cite{Dvali:1997uq}.
These couplings cause the flat directions to be radiatively lifted. For the
case of mediation by the Electroweak Higgs fields, an explicit expression
for the effective potential of the flat direction has
been derived~\cite{Garbrecht:2006az}.
For the case of large Yukawa couplings, in especial for
directions which involve MSSM fields of the third generation, the radiative
effect dominates over the non-renormalizable and the soft SUSY breaking terms,
such that the VEV along the flat direction during inflation can be predicted.
This type of SUSY breaking
does not occur in $D$-term~\cite{Binetruy:1996xj,Halyo:1996pp} or other types
of inflation, where no direct
renormalizable couplings between the inflaton-waterfall and
the MSSM sector occur. Note also, that this effect is not due to the
background curvature, and indeed, calculating this type of
lifting in flat background is a very good approximation.

The second contribution arises also within $F$-term inflation. When taking
supergravity into account, scalar fields generically have Planck-scale
suppressed couplings to the
non-vanishing $F$-terms, which drive inflation~\cite{Dine:1995kz,Dine:1995uk}.
Within minimal supergravity,
a mass square of $3 H^2$, where $H$ is the Hubble rate, arises during
inflation this way. For non-minimal K\"ahler potentials, the mass square can
take different, in particular also negative, values, but is expected
to be of order $H^2$. This type of mass term can however
be forbidden at tree level by imposing a Heisenberg
symmetry~\cite{Binetruy:1987xj,Gaillard:1995az}.

The third possibility, the lifting by the background curvature, is the
subject of the present paper. Since various degrees of freedom of different
spin, albeit being of equal mass, couple to curvature in a different way,
the mass term in the effective potential is renormalized. For the
effective potential in de Sitter space with a chiral multiplet running
in the loop, this is pointed out in~\cite{Garbrecht:2006df}. Here, we extend
the analysis to a vector multiplet and use this result to determine
the lifting of MSSM flat directions during inflation. We find that
the order $H^2$ contributions to the effective potential
arising from the degrees of freedom which acquire masses by the
super Higgs mechanism cancel. Therefore, the only curvature-induced
lifting terms are mediated by the Yukawa-couplings and
come from chiral supermultiplets in the loop.

The single terms in the effective potential in de Sitter background
are known and are constituted by contributions from the scalar
and spin-$\frac 12$ fermion loop
contributions~\cite{Candelas:1975du,Miao:2006pn,Garbrecht:2006df}
as well as the vector boson
loop~\cite{Shore:1979as,Allen:1983dg}. While all authors agree about
the same generic form of the contributions, differences due to
renormalization conditions~\cite{Miao:2006pn}
or due to minor numerical mistakes in the coefficients occur. While we
believe to know the numerically correct answers for the fermionic and the
scalar case~\cite{Miao:2006pn,Garbrecht:2006df}, a rederivation of
the gauge-boson loop contribution is in order. As a novel feature of the
calculation presented here, we allow for a general gauge fixing parameter
$\xi$. The final result should of course be gauge independent, which
is an important consistency check for our calculation. This way, we reproduce
Allen's Ishikawa's result~\cite{Allen:1983dg,Ishikawa:1983kz},
obtained in Landau gauge, $\xi=0$.

\section{Gauge-Invariant Effective Potential for a Charged Scalar
during Inflation}

In this section, we derive the one-loop effective potential in
de Sitter background for a
charged scalar, which gives masses to gauge bosons {\it via} the
Higgs-mechanism. Effective potentials in de Sitter backgrounds were
first calculated for fermion and scalar loops by Candelas
and Raine~\cite{Candelas:1975du}, with numerical inaccuracies within the
finally quoted results however, which motivated
rederivations~\cite{Garbrecht:2006df,Miao:2006pn}.
The effective potential for a
gauge boson loop was first calculated by Shore~\cite{Shore:1979as}
and later on by Allen and Ishikawa~\cite{Allen:1983dg,Ishikawa:1983kz},
who agree with Shore's overall form but find different
numerical coefficients.
For the construction of renormalization group improved effective potentials
in expanding cosmological backgrounds, see~\cite{Elizalde:1993ee}.

In order to show that  the curved space contributions to the effective
potential do not cancel within SUSY, as the flat space contributions do,
we need reliable results with accurately determined coefficients.
Here, we use techniques for deriving
the gauge boson propagator pioneered by Allen and Jacobson~\cite{Allen:1985wd}
and further
elaborated by Kahya and Woodard~\cite{Kahya:2005kj,Kahya:2006ui}.
Employing in addition
DeWitt's method of WKB or adiabatic expansion~\cite{DeWitt:1975ys},
these allow us to find an approximate
expression for the propagator with a general Faddeev-Popov gauge-fixing
parameter $\xi$, and as a novel feature of our derivation, we do not
impose a particular choice on $\xi$. We note that
an exact solution for the propagator in Landau-Lorenz gauge,
$\xi=0$, has recently been obtained by
Tsamis and Woodard~\cite{Tsamis:2006gj}, and results in other specific gauges
are derived in~\cite{Kahya:2005kj}.
The somewhat complementary topic, the effect of the scalar loops on
photons during inflation, is investigated
in~\cite{Prokopec:2002jn,Prokopec:2002uw,Prokopec:2003bx,Prokopec:2003tm,Prokopec:2006ue}.

In our calculations, we use the conventions
of {\it e.g.}~\cite{Kahya:2005kj,Tsamis:2006gj}.
For de~Sitter space expanding at the Hubble rate $H$, we use conformal
coordinates, such that the scale factor is $a=-1/(H\eta)$, where
$\eta \in ]-\infty,0[$ is conformal time.
The metric tensor reads
\begin{equation}
g_{\mu\nu}=a^2 \eta_{\mu\nu}\,,
\end{equation}
where
\begin{equation}
\eta_{\mu\nu}={\rm diag}(-1,\!\!\underbrace{1,...,1}_{D-1\; {\rm times}}\!)
\nonumber
\end{equation}
is the $D$-dimensional Minkowski metric.
We seek to express two-point functions in terms of one de~Sitter-invariant
distance variable.
Therefore,
we introduce the separation between two coordinate points ($\eta=x^0$)
\begin{equation}
\Delta x^2(x;x^\prime)= -(|\eta-\eta^\prime| -{\rm i} \epsilon)^2
+\sum_{i=1}^D|x^i-{x^\prime}^i|^2
\,,
\end{equation}
where the ${\rm i} \epsilon$-term is appropriately chosen to construct a
Feynman-propagator.
The function $\Delta x^2(x;x^\prime)$ is of course not de~Sitter
invariant, but it can be used to construct the
geodesic distance  $\ell(x;x^\prime)$ as
\begin{eqnarray}
 y = 4\sin^2\Big(\frac12 H\ell\Big) = a(\eta)a(\eta^\prime) H^2 \Delta x^2
\label{y}\,.
\end{eqnarray}
It has proved useful in many calculations to use the function
$y(x;x^\prime)$ rather than the more familiar
geodesic distance in order to express
invariant lengths, and we adapt this practice in the following.
We refer especially to~\cite{Tsamis:2006gj} for an appraisal
of the $y$-basis for the construction of bitensors.

In this setup, we consider the gauge invariant Lagrangian
\begin{eqnarray}
\label{L:GI}
\frac 1{\sqrt{-g}}{\cal L}_{\rm GI}=
-\frac 14 F_{\mu\nu}^a F_{\varrho\sigma}^a g^{\mu\nu}g^{\varrho\sigma}
-\frac 12 \left[D_\mu \Phi \right]^T g^{\mu\nu} D_\nu \Phi
-V(\Phi)\,,
\end{eqnarray}
where
\begin{equation}
D_\mu=\mathbbm{1}\nabla_\mu+{\rm i} g T^a A^a_\mu\,,
\end{equation}
and ${\rm i}T^a$, normalized as ${\rm tr} ({\rm i}T^a)^2=\frac 12$,
are the real antisymmetric
generators of gauge transformations of the field $\Phi$, which is
expressed as a real column vector. The field $\Phi$ is decomposed
into a VEV $v=\langle \Phi \rangle$
and into dynamical degrees of freedom $\phi$ as
\begin{equation}
\Phi=\phi+v\,.
\end{equation}
In order to quantize, we apply the Faddeev-Popov procedure.
For that purpose,
the covariant generalization of the flat-space gauge fixing Lagrangian is
\begin{equation}
\label{L:GF}
\frac 1{\sqrt{-g}}{\cal L}_{\rm GF}=
-\frac 1{2\xi}\left(\nabla^\mu A^a_\mu +\xi g \phi^T {\rm i} T^a v\right)^2
\,,
\end{equation}
and the according ghost Lagrangian
\begin{equation}
\frac 1{\sqrt{-g}}{\cal L}_{\rm FP}=-(\nabla_\mu \eta^{a*})
\left[
\nabla^\mu \eta^a + g f^{abc} \eta^b A^{\mu c}
\right] -\xi \eta^{a*}(M^2)^{ab}\eta^b
-\xi g^2\eta^{a*}\eta^b \phi^T T^b T^a v\,,
\end{equation}
where the $\eta^a$ are the anticommuting $\mathbbm{C}$-number ghost
fields.
We can then derive the Feynman rules from
\begin{equation}
{\cal L}={\cal L}_{\rm GI}+{\cal L}_{\rm GF}+{\cal L}_{\rm FP}\,.
\end{equation}

For the calculation of the one-loop effective potential, it is useful
to write down the expansion of ${\cal L}$ to quadratic order in the dynamical
fields,
\begin{eqnarray}
{\cal L}_{\rm qu.}=\int d^D x \sqrt{-g}
\!\!\!&\Bigg\{&\!\!\!
\frac 12 A_\mu^a \nabla^\nu \nabla_\nu A^{a\mu}
-\frac 12 \left(1-\frac 1\xi \right) A_\mu^a \nabla^\mu \nabla_\nu A^{a\nu}
\nonumber
\\
&-&\!\!\!\frac 12 A^a_\mu R^\mu_{\phantom{\mu}\nu} A^{a\nu}
-\frac 12 A^a_\mu (M^2)^{ab} A^{b\mu}
\nonumber
\\
&-&\!\!\!\frac 12 (\partial_\mu \phi^u)(\partial^\mu \phi^u)
-\frac 12 \phi^p \xi (M^2_{\rm G})^{pq} \phi^q
-\frac 12 \phi^r (M_{\rm H}^2)^{rs} \phi^s
\nonumber
\\
&-\!\!\!&(\partial_\mu \eta^{a*})(\partial^\mu \eta^a)
-\eta^{a*} \xi (M^2)^{ab} \eta^b
\Bigg\}
\label{L:qu}
\,.
\end{eqnarray}
We note that the term proportional to the Ricci tensor arises due to
the fact that the covariant derivatives do not commute.
The mass matrix for the gauge bosons is given by
\begin{equation}
\label{M:ab}
\frac 12 (M^2)^{ab}=g^2 v^T T^a T^b v\,.
\end{equation}
By an appropriate choice of the generators $T^a$, it naturally decomposes
into a zero submatrix for the components $a$, $b$ corresponding to
the unbroken generators satisfying $T^{a,b}v =0$ and a non-zero
submatrix for the broken generators with $T^{a,b}v \not=0$. We denote
the part of $M^2$ restricted to the non-zero block as $\bar M^2$.

The scalar mass matrix $M^2_{\rm H}$ occurring in~(\ref{L:qu})
is given by
\begin{equation}
(M^2_{\rm H})^{rs}=\frac{\partial V}{\partial \phi_r \partial \phi_s}\,.
\end{equation}
According to the Higgs-mechanism, it gives masses to the
physical Higgs fields consisting of the components of $\phi$ which are
perpendicular to $T^a v$.
The remaining degrees of freedom are the Goldstone fields spanned by
the vectors $T^a v$, with mass terms
\begin{equation}
\label{M:Goldstone}
\frac 12 \xi(M^2_{\rm G})^{pq}=\xi g^2 v_l T^a_{pl} T^a_{qk} v_k
\,.
\end{equation}
We also recall the fact that while $M^2$ and $M^2_{\rm G}$ have in
general a different number of zero eigenvalues, their spectrum
of non-zero eigenvalues is identical. In particular, this implies
that ${\rm tr}M^2_{\rm G}={\rm tr} M^2$.

Since we want to obtain the effective potential as a function of the modulus
of a flat direction, which we denote by $w$, it is convenient to express
the mass matrices as
\begin{eqnarray}
\label{M:w}
\frac 12 (\bar M^2)^{ab}\!\!\!&=&\!\!\!\frac 12 w^2(\mu^2)^{ab}\,,
\\
\frac 12 \xi(M^2_{\rm G})^{pq}\!\!\!&=&\!\!\!
\frac 12 w^2 \xi (\mu_{\rm G}^2)^{pq}\,,
\end{eqnarray}
where $w=\sqrt{v^i v^i}$. Note that according to this definition,
$M^2_H$ is independent of $w$.

The effective Lagrangian ${\cal L}_{\rm eff}$
is given by the functional relation
\begin{equation}
{\rm i}\int d^D x {\cal L}_{\rm eff}=
\log\left\{{\cal N}^{-1}
\int {\cal D}[A,\,\phi,\,\eta,\,\eta^*]
{\rm e}^{{\rm i} S[A,\,\phi,\,\eta,\,\eta^*]}
\right\}
\,,
\end{equation}
from which we can derive
\begin{eqnarray}
\frac{\partial}{\partial w^2} \int d^D x {\cal L}_{\rm eff.}
\!\!\!&+&\!\!\!{\rm i}{\cal N}^{-1} \frac{\partial {\cal N}}{\partial w^2}
\nonumber
\\
&&\hskip -3.7cm =-\frac{
\int{\cal D}[A,\,\phi,\,\eta,\,\eta^*]\int d^D x
\left\{
\frac{\mu^{ab}}2 A^a_\mu A^{b\mu}
-\xi \mu^{ab} \eta^{*a} \eta^{b}
+\xi \frac{\mu_{\rm G}^{pq}}2 \phi^p \phi^q
\right\}{\rm e}^{{\rm i} S[A,\,\phi,\,\eta,\,\eta^*]}
}
{\int {\cal D}[A,\,\phi,\,\eta,\,\eta^*]
{\rm e}^{{\rm i} S[A,\,\phi,\,\eta,\,\eta^*]}}
\nonumber
\\
&&\hskip -3.7cm =
\int d^D x \sqrt{-g}
\left\{
-\frac 12 (\mu^2)^{ba} {\rm i} {}_\mu[ {}^a\Delta^b(x,x)]^\mu
+\xi (\mu^2)^{ba} {\rm i} \Delta_\eta^{ab}(x,x)
-\frac 12 \xi (\mu_{\rm G}^2)^{qp} {\rm i} \Delta_{\rm G}^{pq}(x,x)
\right\}.
\end{eqnarray}
Here, the functions ${}_\mu[ {}^a\Delta^b(x,x^\prime)]_\nu$,
$\Delta_\eta^{ab}(x,x^\prime)$
and $\Delta_{\rm G}^{pq}(x,x^\prime)$
denote the Feynman propagators for the gauge,
ghost and Goldstone fields, respectively.
The derivative of the effective action with respect to $w^2$ does not
involve any kinetic terms. We can therefore write
\begin{equation}
\int d^D x V_{\rm gauge}=-\int d^D x {\cal L}_{\rm eff}
\end{equation}
for the effective potential arising from the massive gauge-boson loop.
Here we have dropped the terms involving the normalization ${\cal N}$
as an irrelevant integration constant. Decomposing
$V_{\rm gauge}=V_{A}+ V_{\rm G}+V_\eta$
into its individual contributions, we arrive at
\begin{eqnarray}
\label{DV:A}
\frac{\partial V_{A}}{\partial w^2}\!\!\!&=&\!\!\!
\sqrt{-g}\frac 12 (\mu^2)^{ba}{\rm i}  {}_\mu[\Delta^{ab}(x,x)]^\mu\,,
\\
\label{DV:G}
\frac{\partial V_{\rm G}}{\partial w^2}\!\!\!&=&\!\!\!
\sqrt{-g}\frac 12 \xi (\mu^2_{\rm G})^{qp}{\rm i}  \Delta_G^{pq}(x,x)\,,
\\
\label{DV:eta}
\frac{\partial V_\eta}{\partial w^2}\!\!\!&=&\!\!\!
-\sqrt{-g} \xi (\mu^2)^{ba}{\rm i}  \Delta_\eta^{ab}(x,x)\,.
\end{eqnarray}

The propagators for the Goldstone and ghost fields are of the same form,
since they both
correspond to minimally coupled scalar fields. Let $\xi m_i^2$ be
any of the eigenvalues of $\xi M_{\rm G}^2$ or $\xi \bar M^2$.
In the adiabatic limit, $\xi m_i^2\gg H^2$ for all these eigenvalues,
the scalar propagator in de Sitter
background can be expanded as~\cite{Prokopec:2002jn,Prokopec:2003bx,Prokopec:2002uw,Prokopec:2003tm,Garbrecht:2006df}
\begin{eqnarray}
{\rm i}\Delta_{\rm S}(x,x^\prime)=\frac{H^2}{4 \pi^2}\frac{1}{y}
+\frac{1}{16\pi^2}\left({\cal M}^2 -2 H^2\right)
\log \left(\frac {{\cal M}^2}{H^2}y\right)
+\cdots\,,
\end{eqnarray}
where ${\cal M}^2$ stands for either $\xi \bar M^2$ or $\xi M^2_{\rm G}$.

Let us now derive an according expression for the gauge boson propagator
in a general gauge $\xi$. Being a bitensor, it can be decomposed
into the following components\cite{Allen:1985wd,Kahya:2005kj}:
\begin{equation}
{\rm i} {}_\mu[\Delta^{ab}(y)]_\nu 
= B^{ab}(y) \frac{\partial^2 y}{\partial x^\mu \partial x^{\prime\nu}}
+C^{ab}(y) \frac{\partial y}{\partial x^\mu}
\frac{\partial y}{\partial x^{\prime\nu}}\,.
\end{equation}
The solution has to fulfill the Feynman propagator equation
\begin{eqnarray}
\sqrt{-g}\left[\left(
\nabla^\sigma\nabla_\sigma \delta_\mu^{\phantom{\mu}\kappa}
-\left(1-\frac 1\xi\right)\nabla_\mu\nabla^\kappa
-R_\mu^{\phantom{\mu}\kappa}
\right)\delta^{ac}
-(M^2)^{ac}\delta_\mu^{\phantom{\mu}\kappa}
\right]{\rm i} {}_\kappa[{}^c\Delta^b(x,x^\prime)]_\nu
\nonumber
\\
={\rm i} g_{\mu\nu} \delta^{ab} \delta^D(x-x^\prime)\,.
\label{Eqprop}
\end{eqnarray}
Using the efficient relations by Kahya and Woodard~\cite{Kahya:2005kj},
who choose the Feynman gauge $\xi=1$,
we can easily generalize their equations for
$B(y)$ and $C(y)$ as
\begin{eqnarray}
(4y-y^2) B^{\prime\prime} + (2-y)(D-1) B^\prime
-(2-y) (D-1) C - (4y -y ^2) C^\prime -\frac{M^2}{H^2} B \!\!\!&&
\nonumber
\\
+\frac 1\xi\left[
(2-y)B^\prime-D B+(4y-y^2)C^\prime+(2-y)(D+1) C
\right]
\!\!\!&=&\!\!\! 0
\label{EqB}\,,
\\
(2-y)C^\prime - (D-1) C - (2-y)B^{\prime\prime}+(D-1) B^\prime
-\frac{M^2}{H^2} C
\!\!\!&&
\nonumber
\\
+\frac 1\xi \left[
(2-y)B^{\prime\prime}-(D+1) B^\prime + (4y-y^2) C^{\prime\prime}
+(2-y)(D+3)C^\prime - (D+1) C
\right]
\!\!\!&=&\!\!\! 0\label{EqC}\,,
\end{eqnarray}
where the prime denotes a derivative with respect to $y$.

Next, we set $D=4$ and make the {\it ansatz} for the components corresponding
to massive gauge bosons
\begin{eqnarray}
B(y)\!\!\!&=&\!\!\! \beta_1 \frac 1y + \beta_2 
\log \left( \frac{\bar M^2}{H^2} y \right)
+\cdots\,,
\label{Bexp}
\\
C(y)\!\!\!&=&\!\!\! \gamma_1 \frac 1{y^2} +\gamma_2 \frac 1y 
+\gamma_3 \frac 1y \log \left( \frac{\bar M^2}{H^2} y \right) +\cdots\,,
\label{Cexp}
\end{eqnarray}
implementing the DeWitt~\cite{DeWitt:1975ys} adiabatic
expansion of the propagator,
which is applicable when the condition $m_i^2 \gg H^2$
holds, where $m_i^2$ stands again for the the eigenvalues of
$\bar M^2$.
This series {\it ansatz}, which is similarly also applicable
for the scalar and fermion propagators, is to be compared with the alternative
possibility of finding exact solutions for the propagators. In de~Sitter
background, the exact solutions are typically given in terms of hypergeometric
functions, and for the gauge boson propagator, the solution has only been
constructed in particular gauges so far~\cite{Tsamis:2006gj,Kahya:2005kj}.
The expansion {\it ansatz} in contrast appears simpler and works in all
kind of backgrounds.
Its main shortcoming is however, that in the case where the adiabatic
condition $m_i^2 \gg H^2$ does not hold, there may be a constant contribution
to the propagator of order $H^4/m_i^2$, as it is the case for a minimally
coupled scalar field~\cite{Prokopec:2003tm},
which is not reliably captured in adiabatic expansion.
Besides that exact solutions are desirable as a matter of principle,
this justifies the efforts to obtain them.

Plugging the series~(\ref{Bexp}) and~(\ref{Cexp}) into
equations~(\ref{EqB}) and~(\ref{EqC}) and likewise
expanding up to order $y^{-2}$ and $y^{-2} \log y$, we find a system of
equations, that closes for those coefficients which are
explicitly given in~(\ref{Bexp}) and~(\ref{Cexp}).
The result is
\begin{eqnarray}
B(y)\!\!\!&=&\!\!\!
-\frac 1{16\pi^2}(1+\xi) \frac 1y
+\frac{1}{128 \pi^2} \left(
- 3 \frac{\bar M^2}{H^2} - 6 - \frac{\bar M^2}{H^2} \xi^2 + 2\xi 
\right)\log \left(\frac{\bar M^2}{H^2} y \right) +\cdots\,,
\\
C(y)\!\!\!&=&\!\!\!
-\frac 1{16\pi^2}(\xi -1) \frac 1{y^2} +\cdots\,.
\end{eqnarray}
The normalization has been found by expressing the $\delta$-function
in~(\ref{Eqprop}) as given in~\cite{Onemli:2002hr,Garbrecht:2006df}.

In order to feed this result into equation~(\ref{DV:A}), we need to
take the coincidence limit. We therefore note
\begin{eqnarray}
\eta^{\mu\nu} \frac{\partial^2y}{\partial x^\mu \partial x^{\prime\nu}}
\!\!\!&\xrightarrow{y \to 0}&\!\!\! -8 H^2 a^2\,,
\\
\eta^{\mu\nu} \frac 1y
\frac{\partial y}{\partial x^\mu} \frac{\partial y}{\partial x^{\prime \nu}}
\!\!\!&\xrightarrow{y \to 0}&\!\!\! 4 H^2 a^2\,.
\end{eqnarray}
We furthermore introduce a constant physical-length $\varrho$, which
serves as a cutoff scale and is used to regulate the divergent terms
as $y \to 0$~\cite{Garbrecht:2006df}.
The Minkowski trace over the coincident massive
gauge boson propagator then gives
\begin{equation}
{}_\mu\!\! {}^{a}\Delta^{b\mu}(x,x)
=\left[
\frac 1{\varrho^2} \frac {1}{4\pi^2}\left(3 + \xi \right)
+\frac{1}{16\pi^2}\left(3 \bar M^2 +6 H^2 + \xi^2 \bar M^2  -2 \xi H^2 \right)
\log (\varrho^2 \bar M^2)
\right]^{ab}\,,
\end{equation}
while for the scalar propagators, we find the
expressions~\cite{Garbrecht:2006df}
\begin{eqnarray}
\xi\, {}^{a\!}\Delta_{\rm G}^{b}(x,x)\!\!\!&=&\!\!\!\left[
\frac 1{\varrho^2} \frac{1}{4\pi^2} \xi
+\frac{1}{16\pi^2}\left(\xi^2 M_{\rm G}^2 - 2 \xi H^2\right)
\log \left(\varrho^2 \xi M_{\rm G}^2\right)
\right]^{ab}\,,
\\
\xi\, {}^{a\!}\Delta_{\eta}^{b}(x,x)\!\!\!&=&\!\!\!\left[
\frac 1{\varrho^2} \frac{1}{4\pi^2} \xi
+\frac{1}{16\pi^2}\left(\xi^2 \bar M^2 - 2 \xi H^2\right)
\log \left(\varrho^2 \xi \bar M^2\right)
\right]^{ab}
\,.
\end{eqnarray}

We are now ready to calculate the individual contributions to the
effective potential from~(\ref{DV:A}), (\ref{DV:G}) and ~(\ref{DV:eta})
\begin{eqnarray}
V_A\!\!\!&=&\!\!\!{\rm tr}\left[
\frac{M^2}{8\pi^2 \varrho^2}(3+\xi)+ \frac{1}{64 \pi^2}
\left(
3 M^4 + 12 H^2 M^2 + \xi^2 M^4 -4 H^2 \xi M^2
\right)
\log\left(\varrho^2 M^2\right)
\right]\,,
\\
V_{\rm G}\!\!\!&=&\!\!\!{\rm tr}\left[
\frac{M_{\rm G}^2}{8\pi^2 \varrho^2}\xi+ \frac{1}{64 \pi^2}
\left(
\xi^2 M_{\rm G}^4 -4 H^2 \xi M_{\rm G}^2
\right)
\log\left(\varrho^2 M_{\rm G}^2\right)
\right]\,,
\\
V_\eta\!\!\!&=&\!\!\!{\rm tr}\left[
-\frac{M^2}{4\pi^2 \varrho^2}\xi - \frac{1}{64 \pi^2}
\left(
2 \xi^2 M^4 - 8 H^2 \xi M^2
\right)
\log\left(\varrho^2 M^2\right)
\right]\,,
\end{eqnarray}
where we have dropped the finite, analytical, renormalization scheme
dependent terms which are quoted for the scalar and fermion loops
in~\cite{Garbrecht:2006df}.
Using ${\rm tr}M^2_{\rm G}={\rm tr} \bar M^2={\rm tr} M^2$, we
finally find the net result
\begin{eqnarray}
\label{V:gauge}
V_{\rm gauge}=V_A+V_{\rm G}+V_\eta={\rm tr}\left[
\frac{3 M^2}{8\pi^2 \varrho^2} + \frac{1}{64 \pi^2}
\left(
3 M^4 + 12 H^2 M^2
\right)
\log\left(\varrho^2 M^2\right)
\right]\,,
\end{eqnarray}
which agrees with the one by Allen and
Ishikawa~\cite{Allen:1983dg,Ishikawa:1983kz}, which is derived
in Landau gauge, $\xi=0$.
The dependences on the gauge fixing parameter all cancel,
as they should. This is an explicit demonstration of the gauge-independence
of the Faddeev-Popov method in a curved background and also an important
consistency check of the result.
We note that to this end, our observation of gauge invariance is
limited to the parametric range, where the adiabatic condition
$m_i^2\gg H^2$, $\xi m_i^2 \gg H^2$ holds.
In the opposite limit, the situation
may be different, and we have also not given a proof that the gauge-invariance
persists in the limit of Landau-Lorenz gauge, $\xi \to 0$.
We do not consider these questions here,
because we do not have
an exact expression or an appropriate approximation
for the gauge boson propagator in a general gauge $\xi$.
Since, unlike the gauge boson, the scalar field allows for
an adjustable curvature
coupling, which is inherited by the Goldstone boson, whereas the ghost
field is always minimally coupled, it will be interesting
also to investigate the non-adiabatic limit in the future. Gauge invariance
should not be expected in the case of a non-vanishing curvature coupling,
similar to the breakdown of gauge invariance when the Higgs
field in flat space is not settled at its minimum, see
{\it e.g.}~\cite{Binosi:2005yk} for a discussion.

\section{Lifting of Flat Directions}

We now apply the above result to the possible curvature-induced lifting of
flat directions in a spontaneously broken SUSY-gauge theory
during inflation.
In order to complete the spectrum of particles contributing to the one-loop
effective potential, we need to enumerate the additional
scalar and fermionic degrees of freedom which attain masses due to the VEV
along the flat direction.

As we have worked with real representations so far,
it is convenient to rewrite
the complex scalar fields occurring in SUSY in terms of the
standard ${\rm SO}(2)$ representation of complex numbers,
\begin{equation}
a{\rm e}^{{\rm i}\vartheta}
\to a\left(
\begin{array}{cc}
\cos \vartheta & - \sin \vartheta \\
\sin \vartheta & \cos \vartheta
\end{array}
\right)
\,,\quad a,\,\vartheta \in \mathbbm{R}
\,.
\end{equation}
Then, there is for the scalars an additional $D$-term contribution
to the Lagrangian,
\begin{equation}
{\cal L} \supset -g^2  (\phi^T T^a \phi)^2\,.
\end{equation}
When comparing with
the Goldstone boson mass term arising from the gauge-fixing
Lagrangian~(\ref{L:GF}), there is an additional factor ${\rm -i}$ multiplying
$T^a$. This implies that the degrees of freedom, which pick up
masses through the $D$-terms are linear independent
from the Goldstone bosons. Up to
a factor of $1/\xi$, the mass spectrum is however identical to 
the one of the Goldstone Bosons,
and we find the mass matrix~({\it cf.}~(\ref{M:Goldstone}))
\begin{equation}
\frac 12 (M^2_D)^{pq}=g^2 v_l 
\left[ 
\left(
\begin{array}{cc}
0 & -1 \\
1 & 0
\end{array}
\right)
T^a\right]_{pl} 
\left[
\left(
\begin{array}{cc}
0 & -1 \\
1 & 0
\end{array}
\right)
T^a
\right]_{qk} v_k
\,,
\end{equation}
which implies as contribution to the effective
potential~\cite{Garbrecht:2006df}
\begin{equation}
\label{V:D}
V_D={\rm tr}\left[
\frac{M^2}{8\pi^2 \varrho^2}+ \frac{1}{64 \pi^2}
\left(
M^4 -4 H^2 M_{\rm G}^2
\right)
\log\left(\varrho^2 M^2\right)
\right]\,.
\end{equation}

Finally, fermions acquire masses {\it via} the terms
\begin{equation}
{\cal L}\supset-\sqrt g \sqrt 2
\left[
(\phi^\dagger t^a \psi) \lambda^a
+\lambda^{a\dagger} (\psi^\dagger t^a \phi)
\right]\,,
\end{equation}
where we have denoted the fermion multiplet
associated with $\phi$ by $\psi$ and the gauginos by $\lambda^a$.
Supersymmetry or more specific the super-Higgs mechanism ensures that
the square of the resulting fermion mass matrix
$M_\psi^2$ has the same spectrum of
eigenvalues as $M^2$ or $M^2_D$, in particular implying
${\rm tr} M_\psi^2={\rm tr}M^2$. The fermionic contribution
to the effective potential then
is~\cite{Garbrecht:2006df,Miao:2006pn,Candelas:1975du}
\begin{equation}
\label{V:Psi}
V_\psi={\rm tr}\left[
-\frac{M^4}{2\pi^2}\frac{1}{\varrho^2}
+\frac{1}{16\pi^2}
\left(
-M^4-2 H^2 M^2
\right)\log\left(\varrho^2 M^2\right)
\right]\,.
\end{equation}

Summing over the contributions from the gauge boson~(\ref{V:gauge}),
scalar~(\ref{V:D}) and fermion~(\ref{V:Psi})
loops gives for the curvature-induced
effective potential along the flat direction
\begin{equation}
\label{V:Phi}
V_{\rm Higgs}=V_{\rm gauge}+V_D+V_\psi= 0\,,
\end{equation}
up to possible corrections of order $H^4$. This cancellation
for the supersymmetric Higgs mechanism urges
the question whether it was to be expected.
In contrast, the effective potential for two chiral supermultiplets,
consisting together of one massive Dirac fermion and four real
scalars of the mass $m$, is given by~\cite{Garbrecht:2006df}
\begin{equation}
\label{V:chiral}
V_{\rm chiral}=-\frac{3}{8\pi^2} H^2 m^2 \log \left(\varrho^2 m^2 \right)\,,
\end{equation}
which implies that SUSY is broken during inflation and that the effective
potential is indeed lifted. We leave aside here
the interesting question whether the
disappointing cancellation~(\ref{V:Phi}) could have been predicted
in a way which is more than adding the individual contributions to the
effective potential. This leaves also open whether
the cancellation is just accidental or whether it persists beyond
the order $H^2$.

To summarize, we have found that to order $H^2$, there is no curvature-induced
lifting mediated by the  gauge coupling $g$. However, the degrees of freedom
which attain masses {\it via} Yukawa couplings lift the flat direction
as described by the potential $ V_{\rm chiral}$. In general,
$m$ is matrix-valued, and the trace of~(\ref{V:chiral}) is to be taken.
In order to obtain good estimates however, it is in most of the cases
sufficient to replace $m$ by $h w$, where $h$ is the largest Yukawa coupling
of the fields composing the flat direction,
because of the hierarchical nature of the couplings.
We further note that in principle the parameter $\varrho$ needs
to be fixed by a renormalization condition. There is however
no experimental information available which would fix the coefficient
of the operator $m^2 H^2$ at a certain scale. Therefore,
we can only make the statement that a logarithmically running
mass square of order $9/(4\pi^2) h^2 H^2$ is inevitably induced
for flat directions during inflation. We can now
relate these observations to some aspects of flat direction cosmology.

First, we note that $9/16 H^2$ corresponds to a critical value
for a scalar mass square term, since for lower masses, the motion of the
scalar field is overdamped, whereas for larger masses the field
performs damped oscillations around its minimum~\cite{McDonald:1999nc}.
Depending on the value of $\varrho$, a mass term of this size
is likely to be induced by~(\ref{V:chiral}) for those directions
involving top-quark Yukawa couplings, such that $h$ is of order
one. These are directions containing the Higgs field $H_u$ and 
those which contain left- or right handed stops. For directions
which leave the flavour symmetry partially intact at tree level,
this indicates
that depending on the sign  of the induced mass square, the fields
tend to align either parallel or perpendicular to the stop-direction.
If the direction does not involve the top-quark Yukawa coupling,
we can estimate $h$ from the second generation couplings to be of
order $10^{-3}$, such that a much smaller lifting results.

For comparison,
within $F$-term inflation and minimal supergravity, a tree level
mass square of $3 H^2$ is induced~\cite{Dine:1995kz,Dine:1995uk},
which itself is subject to
renormalization group running~\cite{Allahverdi:2001is}.
When assuming a non-minimal K\"ahler potential, the mass term
is $c H^2$, where $c$ is a constant of order one and can be either negative
or positive. In that case, unless the form of the
K\"ahler potential is predicted
by an underlying theory, the curvature induced lifting is just a
correction, which can be absorbed within the unknown parameter $c$.

The case is different when a tree-level mass term of Hubble scale
is absent.
Examples are $D$-term inflation~\cite{Binetruy:1996xj,Halyo:1996pp}
or $F$-term inflation, when endowed with a Heisenberg symmetry which protects
the flatness against supergravity
corrections~\cite{Binetruy:1987xj}, provided that
also effects from spontaneous SUSY breaking are
absent~\cite{Garbrecht:2006az}. For these models, it assumed that
the main contribution to the lifting during inflation comes from
non-renormalizable operators, which only become important at the Planck- or
superstring scale~\cite{Gaillard:1995az,Kolda:1998kc,Enqvist:1998ds}.
In the light of the
present work, which suggests that a logarithmically running
mass of order of the Hubble rate nonetheless arises due to curvature effects,
provided the top-quark Yukawa coupling is involved,
this view needs change.

\section{Summary and Conclusions}

In this work, we rederive Allen's and Ishikawa's
expression~\cite{Allen:1983dg,Ishikawa:1983kz}
for the one-loop,
gauge-boson induced effective potential in de Sitter space. As
a novel feature, the calculation explicitly shows the independence
of the gauge-fixing parameter $\xi$, which is an important consistency
check. The gauge invariance holds in the adiabatic limit, that is
when the gauge boson mass is larger than the Hubble rate. The general
case is left for future investigation.

The gauge boson induced potential can be assembled with the scalar
and fermionic contributions to an effective potential for a SUSY-theory
in de~Sitter background. The cancellation of the flat-space contributions
by virtue of SUSY is another check of the results. Additionally,
we also find a cancellation for the order $H^2$ corrections. Since
such a cancellation is absent for chiral multiplets, it is not necessarily
to be expected. A more detailed understanding of this
observation is desriable.

When applied to MSSM flat directions, our results imply
that gauge-coupling mediated lifting in the inflationary background is absent.
There is however a lifting mediated by Yukawa-couplings.
For directions interacting {\it via} top-quark couplings,
we find that these generically
acquire a logarithmically running mass term, which is of order of the Hubble
rate. For scenarios with non-vanishing $F$-term, this is an additional
and equally important
contribution to supergravity corrections of the same form. For $D$-term models
or $F$-term variants where the flat directions are protected from
supergravity-corrections by additional symmetries, this means that Hubble-scale
mass terms, which are absent at tree-level, may nonetheless be induced at
one-loop in de~Sitter background.
If it turns out that MSSM flat directions indeed acquire VEVs in the
early Universe and that this leads to observable consequences, the
curvature-induced lifting is therefore of importance. Either, observations
can be used to determine the unknown renormalization parameter $\varrho$.
Or, if fundamental theories are developed, which make predictions
for the renormalization conditions, the lifting in inflationary
background needs to be taken into account to test their correctness.

\section*{\large Acknowledgement}
The author is grateful to Apostolos~Pilaftsis for interesting
comments on gauge invariance of effective potentials.

\appendix

\end{document}